\documentclass[english,notitlepage, twocolumn]{revtex4-1}
\usepackage[T1]{fontenc}
\usepackage[latin9]{inputenc}
\usepackage{color}
\usepackage{amsmath}
\usepackage{amssymb}
\usepackage{graphicx}

\usepackage{babel}
\begin{document}

\title{Photon bunching in a rotating reference frame}

\author{Sara Restuccia$^{\dagger,1}$, Marko Toro\v{s}$^{\dagger,2}$,  Graham M. Gibson$^{1}$, Hendrik Ulbricht$^{2}$, Daniele Faccio$^{1}$, Miles J. Padgett$^1$}

\affiliation{$^{1}$School of Physics and Astronomy, University of Glasgow, Glasgow, G12 8QQ, UK}
\affiliation{$^{2}$Department of Physics and Astronomy, University of Southampton, SO17 1BJ, UK}
\email{ miles.padgett@glasgow.ac.uk, daniele.faccio@glasgow.ac.uk \\ $^{\dagger}$These authors contributed equally.}

\begin{abstract}
Although quantum physics is well understood in inertial reference frames (flat spacetime), a current challenge is the search for experimental evidence of non-trivial or unexpected behaviour of quantum systems in non-inertial frames.
Here, we present a novel test of quantum mechanics
in a non-inertial reference frame: we consider Hong--Ou--Mandel (HOM) interference on a rotating platform and study the effect of uniform rotation on the distinguishability of the photons. Both theory and experiments show that the rotational motion induces a relative delay in the photon arrival times at the exit beamsplitter and that this delay is observed as a shift in the position of the HOM dip. This experiment can be extended to a full general relativistic test of quantum physics using satellites in Earth orbit and indicates a new route towards the use of photonic technologies for investigating quantum mechanics at the interface with relativity.
\end{abstract}
\maketitle

{\bf{Introduction.}}
Quantum mechanics and relativity, two cornerstones of modern physics, are separately well formalised and  tested to great precision. Yet the reconciliation of the two in a unified theory has been, and  remains, one of the open problems of theoretical physics. The difficulty in tackling this problem is twofold. First, the mathematical frameworks of these two theories, as well as their conceptual underpinnings, seem to be incompatible with each other~\cite{penrose1996gravity}. Secondly, exploring regimes where both quantum mechanical and relativistic effects are  important has proven to be a significant experimental challenge \cite{hossenfelder2017experimental}. 

These difficulties have led to the pursuit of several approaches~\cite{oriti2009approaches}. Particularly noteworthy is the framework of quantum field theory in curved space--time~\cite{wald1994quantum,brunetti2015advances}, in which some far-reaching progress has been made, with the best  known example being  Hawking's prediction that black holes emit black-body radiation~\cite{Hawking1,hawking1976black}. 

In view of the experimental challenges, several phenomenological models~\cite{hossenfelder2017experimental} have also been proposed, such as minimal length scale~\cite{hossenfelder2013minimal} and semiclassical models ~\cite{DIOSI1984199,ruffini1969systems,kafri2014classical,
diosi1989models,tilloy2016sourcing,gasbarri2017gravity}. Although these models have the advantage of being  testable, they still pose a significant experimental challenge~\cite{kaltenbaek2016macroscopic}.

Quantum mechanical experiments that probe gravity can be divided into (i) local and (ii) non-local classes. The former class probe special relativistic, non-inertial effects in Minkowski space--time, where the Riemann tensor corrections can be neglected, whereas the latter class, which have not yet been experimentally realised, would probe genuinely general relativistic effects, with the outcome
of the experiment being related to the curvature of space--time. So far there have been two experiments of type (i): neutron and atom interferometry
\cite{colella1975observation,nesvizhevsky2002quantum,fixler2007atom},
and entanglement witness of photon pairs \cite{fink2017experimental}.
Both of these experimental classes can be seen, using the equivalence principle, as being implemented either in a uniform gravitational field or in Rindler space--time.
There have been only a handful of proposals of type (ii), all of which are based on the same idea, namely that of one or more small masses interacting
gravitationally~\cite{page1981indirect,anastopoulos2015probing,1742-6596-701-1-012015,krisnanda2017revealing}.
The difficulty of type (ii) experiments can be appreciated by looking at the current experimental limits on the detection of weak gravitational
forces~\cite{ritter1990experimental,schmole2016micromechanical}.
These limits have led the community to also consider an indirect test, where the gravitational interaction, cumulatively over an interval of time, generates an
entangled state~\cite{bose2017spin,marletto2017gravitationally}.

In this letter, we report a novel experiment of type (i) aimed at probing the behaviour of entangled photons in a non-inertial reference frame. Specifically, we demonstrate Hong--Ou--Mandel (HOM) interference ~\cite{hong1987measurement} within a rotating reference frame.  

\begin{figure}
{\includegraphics[width=8cm]{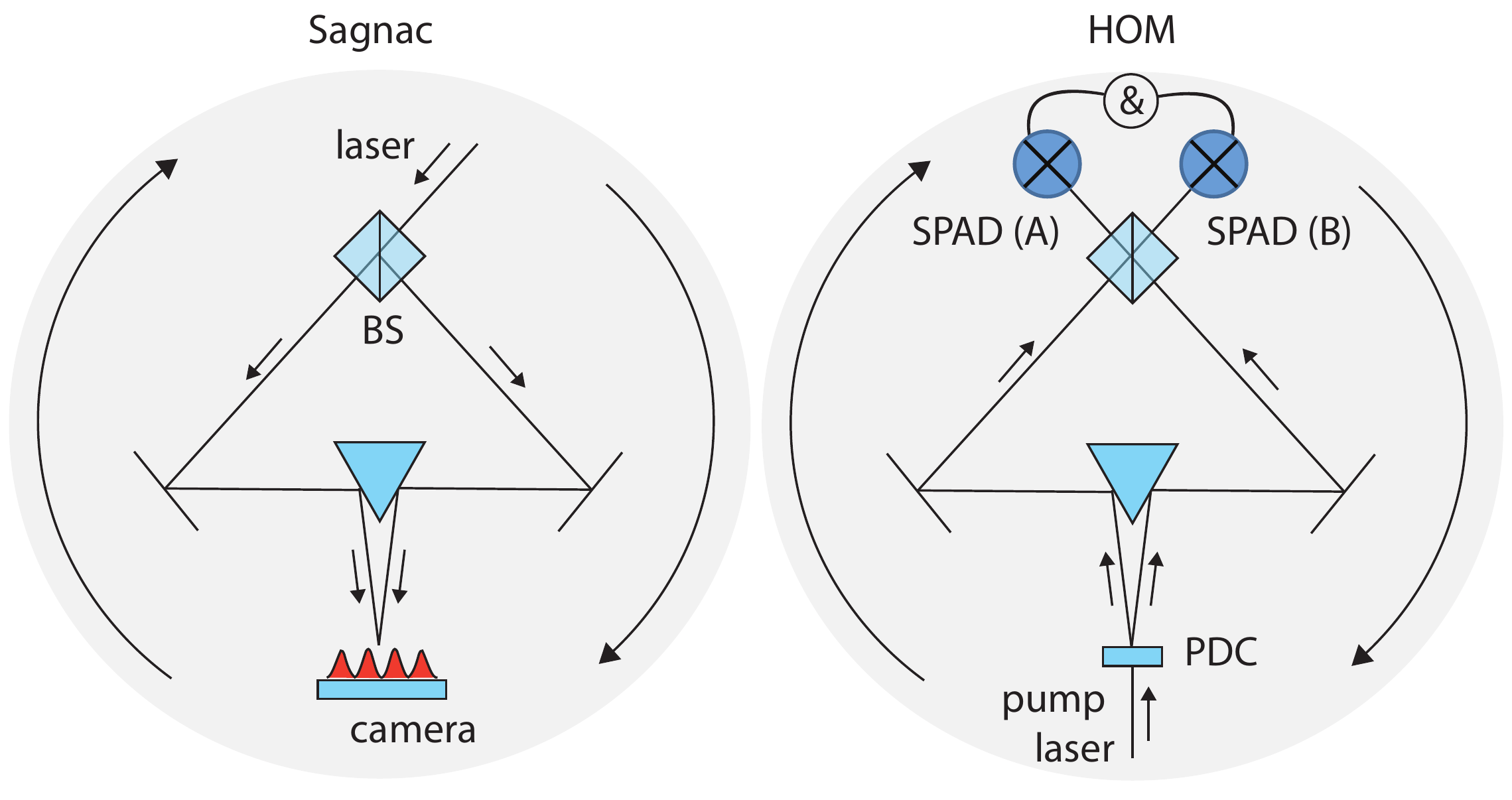}}
\caption{\label{Fig1} Rotating interferometer schematic layouts: `Sagnac' indicates the classical interferometer formed from two counter-propagating beams that then overlap at a small angle on a camera that measures the resulting interference fringes. Platform rotation is detected as a lateral shift in the fringes. `HOM' indicates the quantum Hong--Ou--Mandel interferometer that can be implemented on the same platform as the classical Sagnac interferometer by introducing a down-conversion crystal in the plane of the camera. The two down-converted photons enter the interferometer and propagate in opposite directions, interfere at the beamsplitter (BS), and are coincidence-counted at the two BS output ports with single-photon avalanche detectors, SPADs (A) and (B). Rotation is detected by measuring a variation in the coincidence counts.}
\end{figure}

Key to our experiments is the recognition that a HOM interference experiment is similar in  configuration to a Sagnac interferometer. The latter is based on a ring cavity in which light coupled to the clockwise and anticlockwise paths interferes at the output.  It is well known that  rotation of the frame results in a phase shift in the interference and this effect is the basis of an optical gyroscope ~\cite{sagnac1913ether,sagnac1913preuve}. The HOM setup may also take the form of a ring cavity, but this time one in which the down-converted photons are created within the cavity and take clockwise and anticlockwise paths to interfere at the output. As mentioned above, rotation induces a phase shift between the clockwise and anticlockwise beams that is observed at the output of a generic interferometer  (see Fig.~\ref{Fig1}). The question is whether the HOM dip undergoes a corresponding change. By measuring such a shift within a rotating frame, we obtain a relationship between the classical phase shift and the HOM path-length shift. 
In this way, we combine, in a single experiment, the relativistic Sagnac effect with a quantum mechanical HOM effect. This experiment also has the potential for generalisation to a full general relativistic test of quantum physics using satellites in Earth orbit, based on the gravitomagnetic clock effect~\cite{cohen1993standard}.\\

{\bf{Theory.}}
The rotating HOM interferometer is  shown schematically in Fig.~\ref{Fig1}. A photon source generates a pair of indistinguishable photons that are separated from each other using a knife-edge prism and sent to the clockwise and anticlockwise paths within the interferometer. These photons are then recombined at a beamsplitter. Whenever two indistinguishable photons combine at the beamsplitter at the same time, quantum interference dictates that they will bunch and therefore always exit from the same output port (although which  port this  will be is completely random and cannot be known a priori). Because both photons are emitted from the same port, the detectors see a large decrease in coincidence counts compared with when the photons arrive at different times (when they do not bunch).  

The experimental setup in Fig.~\ref{Fig1} can be modelled using the theoretical description of the HOM interferometer~\cite{ou1988observation}, where the rotational motion of the platform generates a time delay in the arrival times of the two photons. This time delay introduces distinguishability between the two photons, which in turn modifies the interference at the output. We can estimate the time delay $\Delta t$ using the formula for Sagnac delay (which is a classical relativistic effect)  at small rotational frequencies~\cite{sagnac1913ether,sagnac1913preuve,gourgoulhon2016special}:
\begin{equation}
\Delta t=8\pi\frac{Af}{c^{2}},\label{eq:sagnac}
\end{equation}
where $A$ is the area enclosed by the trajectories of the two photons and $f$ is the rotational frequency of the frame.
The expression in equation~(\ref{eq:sagnac}) is the same for both the inertial, laboratory observer and  the non-inertial, co-rotating observer attached to the detector \cite{post}: this is a consequence of the non-relativistic motion of matter.

We consider a generic initial two-photon state:
\begin{equation}
|\chi_{\mathrm{i}}\rangle=\int_{0}^{\infty}\mathrm{d}\omega_{1}\int_{0}^{\infty}\mathrm{d}\omega_{2}\,\psi(\omega_{1},\omega_{2})\hat{a}^{\dagger}(\omega_{1})\hat{b}^{\dagger}(\omega_{2})|0\rangle,
\label{chii} 
\end{equation}
where $\hat{a}(\omega_{1})$ and $\hat{b}(\omega_{2})$ denote the
annihilation operators for modes of frequency $\omega_{1}$ propagating
in the clockwise path and for modes of frequency $\omega_{2}$ propagating in
the anticlockwise path, respectively, and $\psi$ is a $\mathbb{C}$-valued
function. We now assume that the two-photon state, before the interaction
with the beamsplitter, evolves to
\begin{equation}
|\chi_{\mathrm{e}}\rangle=\int_{0}^{\infty}\mathrm{d}\omega_{1}\int_{0}^{\infty}\mathrm{d}\omega_{2}\,\psi(\omega_{1},\omega_{2})\mathrm{e}^{-\mathrm{i}\phi}\hat{a}^{\dagger}(\omega_{1})\hat{b}^{\dagger}(\omega_{2})|0\rangle,\label{eq:xi}
\end{equation}
where $\phi$ is the accumulated phase, which depends on the experimental
setup (e.g. on the length of the optical fibre). Specifically, we assume
that the phase is given by
$\phi(\omega_{1},\omega_{2},f)=\omega_{1}\left(t^{(+)}(f)+\delta t_{p}\right)+\omega_{2}t^{(-)}(f)$, %
where $t^{(+)}$ and $t^{(-)}$ denote the times of flight from the source
to the beamsplitter for two classical light signals propagating
in the clockwise and anticlockwise directions, respectively.
$\delta t_{p}$ denotes an additional time delay due to differences in optical path between the two interferometer arms (which can be controlled by fine-tuning the position of one of the coupling optics and can be used to scan the interferometer delay).

We now consider the beamsplitter transformation:
\begin{equation}
\begin{pmatrix}
\hat{a}^{\dagger}\\[4pt]
\hat{b}^{\dagger}
\end{pmatrix}\rightarrow\begin{pmatrix}
\frac{1}{\sqrt{2}}(i\hat{a}^{\dagger}+\hat{b}^{\dagger})\\[4pt]
\frac{1}{\sqrt{2}}(\hat{a}^{\dagger}+i\hat{b}^{\dagger})
\end{pmatrix}. \label{eq:bs}
\end{equation}
Using equations~(\ref{eq:xi}) and (\ref{eq:bs}),  we obtain the two-photon state after the interaction
with the beamsplitter:
\begin{gather}\begin{split}
|\chi_{\text{bs}}\rangle={}& \frac{1}{2}\int_{0}^{\infty}\mathrm{d}\omega_{1}\int_{0}^{\infty}\mathrm{d}\omega_{2}\,\mathrm{e}^{-\mathrm{i}\{\omega_{1}[t^{(+)}(f)+\delta t_{p}]+\omega_{2}t^{(-)}(f)\}} \\
& \times \psi(\omega_{1},\omega_{2})[\mathrm{i}\hat{a}^{\dagger}(\omega_{1})+\hat{b}^{\dagger}(\omega_{2})]\\&\times [\hat{a}^{\dagger}(\omega_{1})+\mathrm{i}\hat{b}^{\dagger}(\omega_{2})]|0\rangle.\label{eq:be}
\end{split}\raisetag{12pt}
\end{gather}
We now consider the coincidence measurement:
\begin{equation}
\hat{\Pi}=\int_{0}^{\infty}\mathrm{d}\omega_{3}\int_{0}^{\infty}\mathrm{d}\omega_{4}\,\hat{a}^{\dagger}(\omega_{3})\hat{b}^{\dagger}(\omega_{4})|0\rangle\langle0|\hat{a}(\omega_{3})\hat{b}(\omega_{4}).\label{eq:PI}
\end{equation}
Using equations~(\ref{eq:be}) and (\ref{eq:PI}), it is then straightforward
to obtain the probability $p(\delta t_{p})=\langle\chi_{\mathrm{bs}}|\Pi|\chi_{\mathrm{bs}}\rangle$
of coincidence events. Specifically, we find
\begin{multline}
p(\delta t_{p})=\frac{1}{2}-\frac{1}{2}\int_{0}^{\infty}  \mathrm{d}\omega_{1}\int_{0}^{\infty}\mathrm{d}\omega_{2}\psi(\omega_{2},\omega_{1})^{*}\psi(\omega_{1},\omega_{2}) \\
 \times \mathrm{e}^{\mathrm{i}(\omega_{2}-\omega_{1})[\Delta t(f)+\delta t_{p}]},\label{eq:hom}
\end{multline}
where we have assumed $\int_{0}^{\infty}\mathrm{d}\omega_{1}\int_{0}^{\infty}\mathrm{d}\omega_{2}|\,\psi(\omega_{1},\omega_{2})|^{2}=1$,
and $\Delta t=t^{(+)}-t^{(-)}$ is the Sagnac delay given by equation~(\ref{eq:sagnac}).
Note that equation~(\ref{eq:hom}) is the usual HOM coincidence probability
but now with a dependence on rotational frequency $f$ in the time
delay $\Delta t(f)$. 
For example, choosing $\psi(\omega_{1},\omega_{2})=g(\omega_{1})g(\omega_{2})$, where $g(\omega)\approx(2\pi\sigma^{2})^{-1/4}\mathrm{e}^{-(\omega-\mu)^{2}/(4\sigma^{2})}$
and $\mu\gg\sigma$, we find a simple expression
\begin{equation}
p(\delta t_{p};f,A,\sigma)=\frac{1}{2}\left(1-\exp\!\!\left[-\frac{\left(\dfrac{8\pi Af}{c^{2}}+\delta t_{p}\right)}{2\left(\dfrac{1}{\sqrt{2}\,\sigma^{2}}\right)}^{2}\right]\right).\label{eq:gauss}
\end{equation}
This result implies a linear relation between the measured delay (the position of the HOM dip in the coincidence count), the area $A$ of the interferometer and the rotation frequency $f$. Moreover, the amplitude of the shift is exactly equal to that expected for a classical Sagnac interferometer. \\

\begin{figure}[t]
\includegraphics[width=7cm]{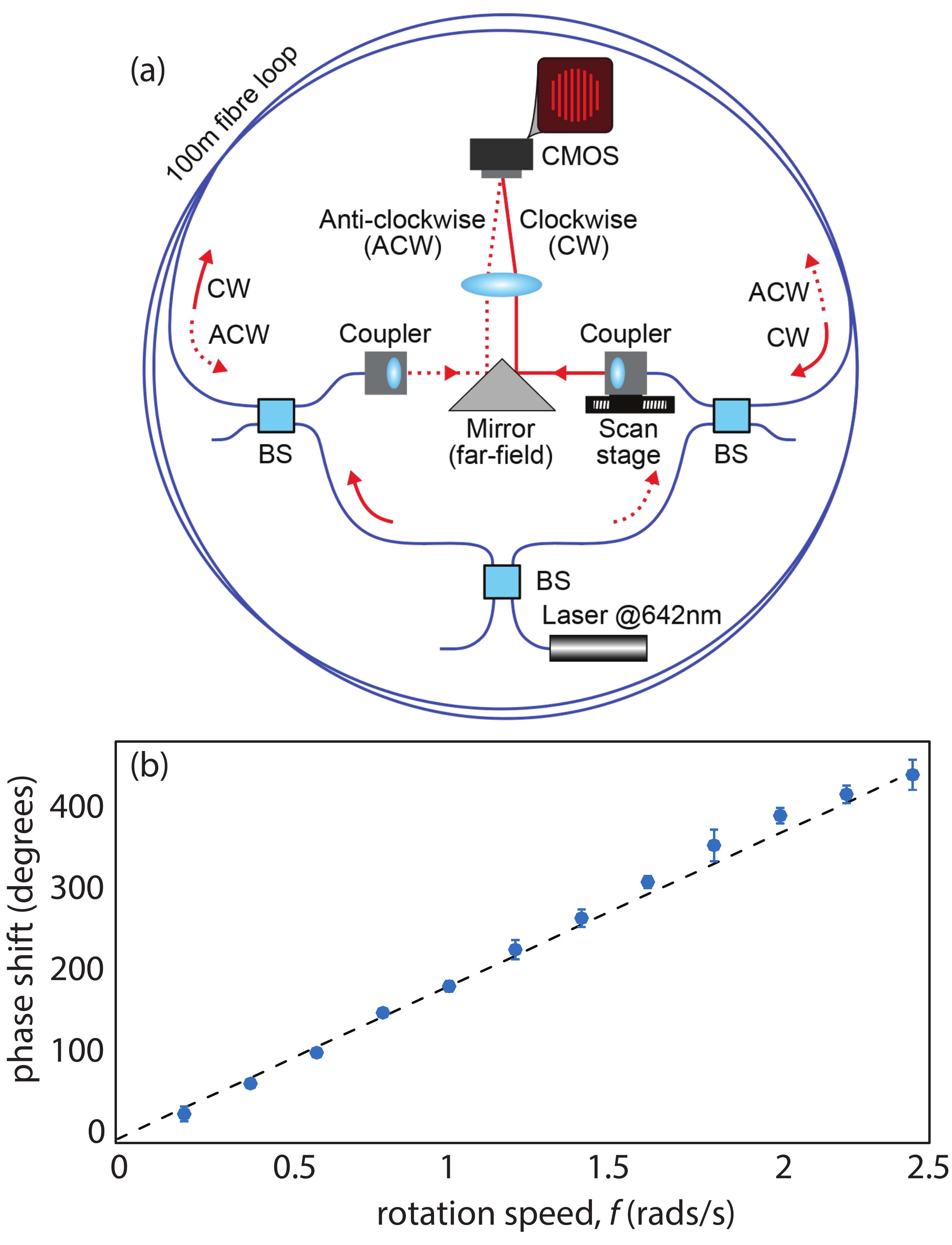}\
\caption{Classical experiment. (a) Experimental layout. (b) Results showing measured Sagnac phase shifts (circles) and fitted line (dashed).
 \label{exp1}}
\end{figure}
\begin{figure}[t]
\includegraphics[width=7cm]{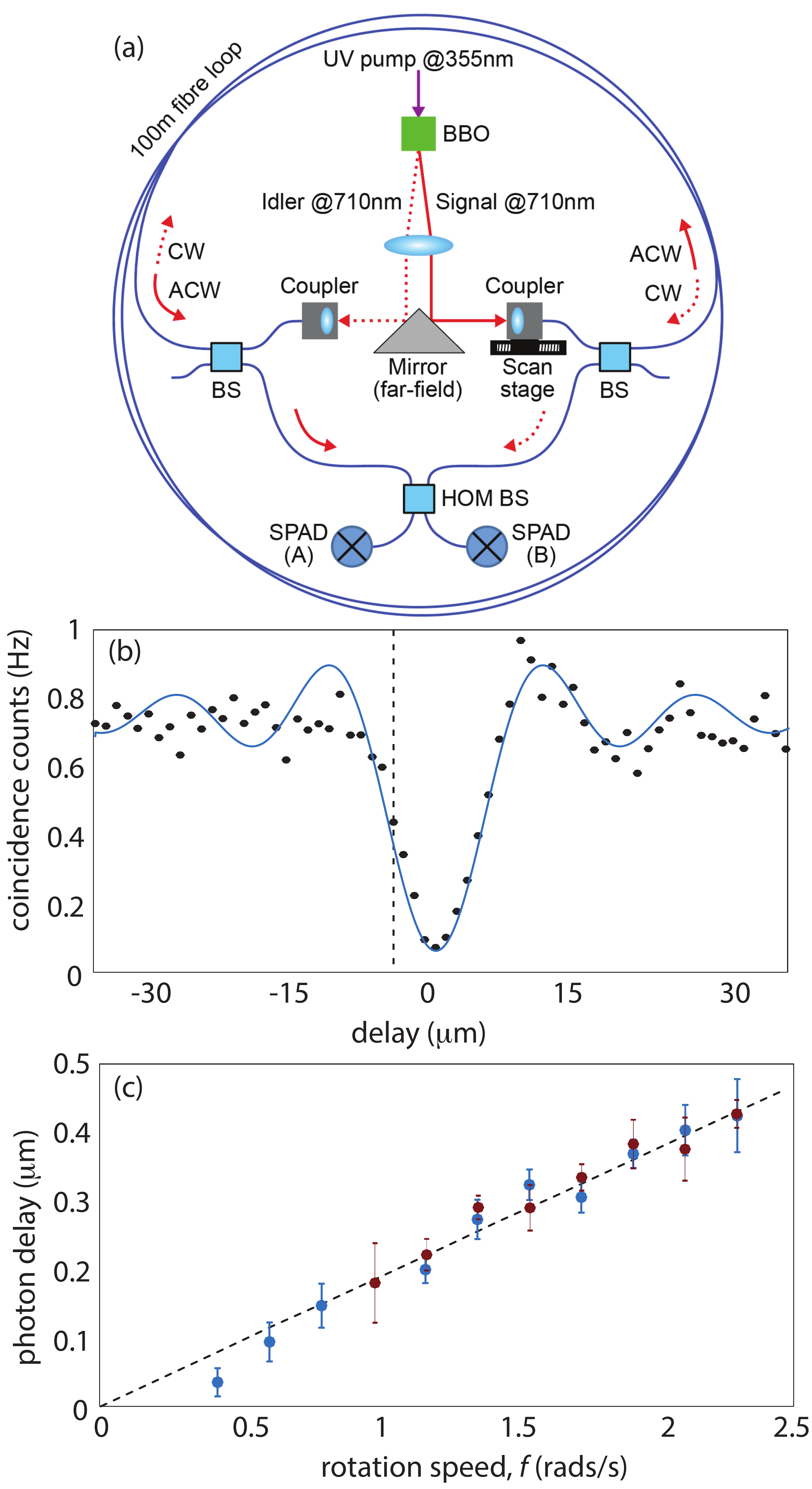}\
\caption{Quantum experiment. (a) Experimental layout. (b) HOM interference dip measured by scanning the delay stage with no rotation of the experiment. The vertical dashed line marks the delay corresponding to the point of maximum steepness of the HOM dip. The shift can then be measured by fixing the stage at this delay and observing the changes in coincidence counts when the interferometer is in rotation. (c) Results showing measured HOM interference shift (circles) and fitted line (dashed).
 \label{exp2}}
\end{figure}

{\bf{Experimental results.}}
Experiments were performed by implementing the scheme shown in Fig.~\ref{Fig1}. 
To create a robust interferometer, rather than using free-space optics to form the clockwise and anticlockwise loops we use a polarisation-maintaining fibre.  The signal and idler beams are separated from each other using a knife-edge prism in the far field of the down-conversion crystal and both photons are subsequently coupled into the fibre, which forms the remainder of the optical system.  A further advantage of a fibre-based approach is that a long length of fibre can be coiled to give many turns and hence increase the sensitivity of the shifts due to the rotation.  This fibre-based approach could have been implemented using separate fibres for the clockwise and anticlockwise loops. However, separate fibres render the interferometer susceptible to drift due to unwanted temperature-based path-length variations, affecting each fibre differently. Here we insert additional beamsplitters after the coupling optics to allow us to use the same fibre for both the clockwise and anticlockwise loops, albeit with a reduction in optical coupling efficiency.

Details of the actual experimental layouts are shown in Figs.~\ref{exp1}(a) and \ref{exp2}(a): the fibre-based interferometer is composed of a 100~m long fibre wound $N=35$ times around a 0.908~m diameter loop. This provides an effective area of $A=N\pi r^2=22.7$~m$^2$. We first performed a classical experiment  to calibrate the setup for the quantum measurements. We coupled a laser (642~nm wavelength) into the interferometer and measured the shifts in the spatial interference fringes on a CMOS camera at the output where the two counter-propagating signals overlap at a small angle. The results are shown in Fig.~\ref{exp1}(b), where we plot the averaged measured phase shifts $\Delta\phi$ together with their standard deviation over 50 measurement runs. We also note that the phase shifts are obtained by performing the average (in absolute value) between a clockwise and an anticlockwise rotation measurement. This is in order to minimise the systematic deviations due to mechanical deformations caused by the presence of centrifugal acceleration ($\sim\!0.3g$). The dashed line shows the best fit to the data with slope $\Delta\phi_\mathrm{fit}=167\pm 4$~deg\,Hz$^{-1}$. This is to be compared with the theoretical estimate based on the standard formula for the Sagnac effect, $\Delta\phi=8\pi A f/(\lambda c)=170$~deg\,Hz$^{-1}$, indicating  good agreement. 

The HOM interference experiments were carried out with the same interferometer but with two modifications. We removed the laser and replaced it with two single-photon avalanche detectors (SPADs), with which we then measured coincidence rates at the output of the HOM beamsplitter. We also replaced the CMOS camera with a down-conversion photon-pair source comprising  a 355~nm pump laser and a barium borate (BBO) crystal. The photon pairs are generated with an angular separation, both retracing the path of the classical setup, are separated by the prism, and interfere at the HOM beamsplitter, as shown in Fig.~\ref{exp2}(a). A delay stage on one of the fibre couplers provides a scan over the temporal delay: a typical result measured in the non-rotating case is shown in Fig.~\ref{exp2}(b), with the typical `HOM dip' (a drop in coincidence rates) when the photons arrive simultaneously at the HOM beamsplitter. 

A rotating HOM experiment involves comparing full scans of the HOM dip and analysing the shift of its position as a function of rotation speed. This approach provides photon path delay measurements with a precision of the order of a micrometre. However, recent work~\cite{attoHOM} has shown that this precision can be greatly increased by fixing the interferometer delay close to the point of maximum steepness of the HOM interference dip, indicated by a vertical dashed line in Fig.~\ref{exp2}(b). Shifts of the dip due to changes in photon path (in our case arising from rotation) are then measured by observing changes in the coincidence counts as the interferometer is rotated. Using the initially acquired HOM dip as a maximum-likelihood estimator, we can map the coincidence counts to photon path delay. This allows us to measure very small photon path delays down to 100~nm and smaller. 

Figure~\ref{exp2}(c) shows the mapped photon delays inferred from changes in coincidence counts,  for increasing rotation rates, with data acquired over different days (blue/red circles). As for the classical experiment, we plot the average delays and standard deviations for 50 separate measurement runs and for the difference between clockwise and anticlockwise rotation. The dashed line shows the best fit to the data with slope $\Delta x_\mathrm{fit}=200\pm 12$~nm\,Hz$^{-1}$. These results clearly show that rotation can, for a fixed physical path length of the interferometer, modify the degree of distinguishability of two entangled photons and thus modify the quantum interference between the two photons. As such, this demonstrates a clear influence of non-inertial motion on non-classical photon states.\\

{\bf{Discussion and conclusions.}}
We can  compare the quantum and the classical results, for example by taking the ratio of the slopes of the two fitted lines in Figs.~\ref{exp1}(c) and \ref{exp2}(c): $\Delta x_\textrm{fit}/\Delta\phi_\textrm{fit}=1.2\pm0.07$~nm\,deg$^{-1}$. Path and phase differences in an interferometer are related by $\Delta x/\Delta \phi=\lambda/2\pi$. If we use the vacuum wavelength for the classical laser (642~nm), we find that $\Delta x_\mathrm{fit}/\Delta\phi_\textrm{fit}$ and $\Delta x/\Delta \phi$ differ by a multiplicative factor of $1.478\pm0.09$.  
This value is compatible with the refractive index $n\sim1.45$ of the optical fibre, as may intuitively  be expected based on the fact that the classical measurement is sensitive to phase (and does not depend on the fibre index $n$~\cite{post}) whereas the quantum measurements rely on time delay (which does depend on $n$).

The experiments proposed here could be generalised to a scenario involving a satellite mission~\cite{gronwald1997gravity}. Specifically, instead of considering the Sagnac effect, we can consider its generalisation to curved space--time~\cite{ruggiero2003relativistic,ruggiero2005sagnac}, i.e. the gravitomagnetic clock effect~\cite{cohen1993standard}. The rotating platform would be replaced by satellites
orbiting around the Earth: two photons would be  transmitted in counter-propagating directions around the Earth, using three or more satellites, back to the satellite of origin, where they are interfered and detected. After a
full revolution around the Earth, the difference in the arrival times of the two photons is given by the formula for the gravitomagnetic effect~\cite{tartaglia2000detection,ciufolini2002time,ciufolini2003gravitomagnetic}:
$\Delta t\sim G J/(R c^{4})$,
where $J$ is the angular momentum of the Earth, $R$ is the distance from its centre and $G$ is the gravitational constant. This relativistic effect would then modify the quantum interference for the two-photon state: the gravitationally induced photon delay would be of the order of $\Delta t \approx 10^{-16}$~s. This delay is only one order of magnitude smaller than recent measurements based on HOM interferometry \cite{attoHOM} and can be increased by increasing the number of revolutions around the Earth. 

The measurement technique presented here builds upon recent developments in quantum sensing and constitutes a new approach to testing the interaction between quantum mechanics and special relativity, with routes towards the inclusion of non-trivial space and space--time curvatures, for example through non-uniform acceleration of the system or extension to gravitational fields with satellite-based experiments.\\

{\bf{Acknowledgements.}}
We acknowledge discussions with Peter Horak and Giulio Gasbarri, and financial support from The Leverhulme Trust, the Foundational Questions
Institute (FQXi) and EPSRC (UK Grants EP/P006078/2 and EP/R030081/1).


\begin{thebibliography}{10}

\bibitem{penrose1996gravity}
Penrose, R.
\newblock On gravity's role in quantum state reduction.
\newblock {\em Gen. Rel. Grav.}, 28(5):581--600, 1996.

\bibitem{hossenfelder2017experimental}
Hossenfelder, S (ed.).
\newblock {\em Experimental Search for Quantum Gravity}.
\newblock Springer, 2018.

\bibitem{oriti2009approaches}
Oriti, D (ed.).
\newblock {\em Approaches to Quantum Gravity: Toward a New Understanding of
  Space, Time and Matter}.
\newblock Cambridge University Press, 2009.

\bibitem{wald1994quantum}
Wald, R. M.
\newblock {\em Quantum Field Theory in Curved Spacetime and Black Hole
  Thermodynamics}.
\newblock University of Chicago Press, 1994.

\bibitem{brunetti2015advances}
Brunetti, R.,  Dappiaggi, C.,  Fredenhagen, K. \& Yngvason, J. (eds.).
\newblock {\em Advances in Algebraic Quantum Field Theory}.
\newblock Springer, 2015.

\bibitem{Hawking1}
Hawking, S. W.
\newblock Black hole explosions?
\newblock {\em Nature}, 248:30--31, 1974.

\bibitem{hawking1976black}
Hawking, S. W.
\newblock Black holes and thermodynamics.
\newblock {\em Phys. Rev. D}, 13(2):191, 1976.

\bibitem{hossenfelder2013minimal}
Hossenfelder, S.
\newblock Minimal length scale scenarios for quantum gravity.
\newblock {\em Living Rev. Rel.}, 16(1):2, 2013.

\bibitem{DIOSI1984199}
Di\'{o}si, L.
\newblock Gravitation and quantum-mechanical localization of macro-objects.
\newblock {\em Phys. Lett. A}, 105(4):199 -- 202, 1984.

\bibitem{ruffini1969systems}
Ruffini R. \& Bonazzola, S.
\newblock Systems of self-gravitating particles in general relativity and the
  concept of an equation of state.
\newblock {\em Phys. Rev.}, 187(5):1767, 1969.

\bibitem{kafri2014classical}
Kafri, D., Taylor, J. M. \& Milburn, G. J.
\newblock A classical channel model for gravitational decoherence.
\newblock {\em New J. Phys.}, 16(6):065020, 2014.

\bibitem{diosi1989models}
Di{\'o}si, L.
\newblock Models for universal reduction of macroscopic quantum fluctuations.
\newblock {\em Phys. Rev. A}, 40(3):1165, 1989.

\bibitem{tilloy2016sourcing}
Tilloy A. \& Di{\'o}si, L.
\newblock Sourcing semiclassical gravity from spontaneously localized quantum
  matter.
\newblock {\em Phys. Rev. D}, 93(2):024026, 2016.

\bibitem{gasbarri2017gravity}
Gasbarri, G.,  Toro{\v{s}}, M.,  Donadi, S. \&  Bassi, A.
\newblock Gravity induced wave function collapse.
\newblock {\em Phys. Rev. D}, 96(10):104013, 2017.

\bibitem{kaltenbaek2016macroscopic}
 Kaltenbaek, R. et~al.
\newblock Macroscopic quantum resonators (maqro): 2015 update.
\newblock {\em EPJ Quantum Technol.}, 3(1):5, 2016.

\bibitem{colella1975observation}
Colella, R., Overhauser, A. W. \& Werner, S. A.
\newblock Observation of gravitationally induced quantum interference.
\newblock {\em Phys. Rev. Lett.}, 34(23):1472, 1975.

\bibitem{nesvizhevsky2002quantum}
Nesvizhevsky, V. V. et~al.
\newblock Quantum states of neutrons in the {Earth's} gravitational field.
\newblock {\em Nature}, 415(6869):297, 2002.

\bibitem{fixler2007atom}
Fixler, J. B., Foster, G. T., McGuirk, J. M. \& Kasevich, M. A.
\newblock Atom interferometer measurement of the {Newtonian} constant of gravity.
\newblock {\em Science}, 315(5808):74--77, 2007.

\bibitem{fink2017experimental}
Fink, M. et al.
\newblock Experimental test of photonic entanglement in accelerated reference
  frames.
\newblock {\em Nat. Commun.}, 8:15304, 2017.

\bibitem{page1981indirect}
Page, D. N. \& Geilker, C. D.
\newblock Indirect evidence for quantum gravity.
\newblock {\em Phys. Rev. Lett.}, 47(14):979, 1981.

\bibitem{anastopoulos2015probing}
Anastopoulos, C. \& Hu, B.-L.
\newblock Probing a gravitational cat state.
\newblock {\em Class. Quantum Grav.}, 32(16):165022, 2015.

\bibitem{1742-6596-701-1-012015}
Derakhshani, M., Anastopoulos, C. \& Hu, B.-L.
\newblock Probing a gravitational cat state: experimental possibilities.
\newblock {\em J. Phys.: Conf. Ser.}, 701(1):012015, 2016.

\bibitem{krisnanda2017revealing}
Krisnanda, T.,  Zuppardo, M.,  Paternostro, M. \&  Paterek, T.
\newblock Revealing nonclassicality of inaccessible objects.
\newblock {\em Phys. Rev. Lett.}, 119(12):120402, 2017.

\bibitem{ritter1990experimental}
Ritter, R. C., Goldblum, C. E., Ni, W.-T., Gillies, G. T. \& 
  Speake, C. C.
\newblock Experimental test of equivalence principle with polarized masses.
\newblock {\em Phys. Rev. D}, 42(4):977, 1990.

\bibitem{schmole2016micromechanical}
Schm{\"o}le, J.,  Dragosits, M.,  Hepach, H. \&  Aspelmeyer, M.
\newblock A micromechanical proof-of-principle experiment for measuring the
  gravitational force of milligram masses.
\newblock {\em Class. Quantum Grav.}, 33(12):125031, 2016.

\bibitem{bose2017spin}
Bose, S. et al.
\newblock Spin entanglement witness for quantum gravity.
\newblock {\em Phys. Rev. Lett.}, 119(24):240401, 2017.

\bibitem{marletto2017gravitationally}
Marletto C. \&  Vedral, V.
\newblock Gravitationally induced entanglement between two massive particles is
  sufficient evidence of quantum effects in gravity.
\newblock {\em Phys. Rev. Lett.}, 119(24):240402, 2017.

\bibitem{hong1987measurement}
Hong, C.-K., Ou, Z.-Y. \&  Mandel, L.
\newblock Measurement of subpicosecond time intervals between two photons by
  interference.
\newblock {\em Phys. Rev. Lett.}, 59(18):2044, 1987.

\bibitem{sagnac1913ether}
Sagnac, G.
\newblock L'{\'e}ther lumineux d{\'e}montr{\'e} par l'effet du vent relatif
  d'{\'e}ther dans un interf{\'e}rom{\`e}tre en rotation uniforme.
\newblock {\em C. R. Acad. Sci.}, 157:708--710, 1913.

\bibitem{sagnac1913preuve}
Sagnac, G.
\newblock Sur la preuve de la r{\'e}alit{\'e} de l'{\'e}ther lumineux par
  l'exp{\'e}rience de l'interf{\'e}rographe tournant.
\newblock {\em C. R. Acad. Sci.}, 157:1410--1413, 1913.

\bibitem{cohen1993standard}
Cohen, J. M. \& Mashhoon, B.
\newblock Standard clocks, interferometry, and gravitomagnetism.
\newblock {\em Phys. Lett. A}, 181(5):353--358, 1993.

\bibitem{ou1988observation}
Ou, Z.-Y. \& Mandel, L.
\newblock Observation of spatial quantum beating with separated photodetectors.
\newblock {\em Phys. Rev. Lett.}, 61(1):54, 1988.

\bibitem{gourgoulhon2016special}
Gourgoulhon, {\'E}.
\newblock {\em Special Relativity in General Frames: From Particles to Astrophysics}.
\newblock Springer, 2016.

\bibitem{post}
Post, E. J.
\newblock Sagnac effect.
\newblock {\em Rev. Mod. Phys.}, 39:475, 1967.

\bibitem{attoHOM}
Lyons, A. et al.
\newblock Attosecond-resolution Hong--Ou--Mandel interferometry.
\newblock {\em Sci. Adv.}, 4:eaap9416, 2018.

\bibitem{gronwald1997gravity}
Gronwald, F., Gruber, E., Lichtenegger, H. \& Puntigam, R. A.
\newblock Gravity Probe C(lock) -- probing the gravitomagnetic field of the Earth
  by means of a clock experiment.
\newblock {Preprint at https://arxiv.org/abs/gr-qc/9712054 (1997)}.

\bibitem{ruggiero2003relativistic}
Ruggiero, M. L. \&  Rizzi, G.
\newblock {The relativistic Sagnac effect: two derivations}.
\newblock In {\em Relativity in Rotating Frames}, eds. Ruggiero, M. L. \&  Rizzi, G.
\newblock Kluwer-Springer, 2003.

\bibitem{ruggiero2005sagnac}
Ruggiero, M. L.
\newblock The Sagnac effect in curved space--times from an analogy with the
  Aharonov--Bohm effect.
\newblock {\em Gen. Rel. Grav.}, 37(11):1845--1855, 2005.

\bibitem{tartaglia2000detection}
Tartaglia, A.
\newblock Detection of the gravitomagnetic clock effect.
\newblock {\em Class. Quantum Grav.}, 17(4):783, 2000.

\bibitem{ciufolini2002time}
Ciufolini I. \& Ricci, F.
\newblock Time delay due to spin inside a rotating shell.
\newblock {\em Class. Quantum Grav.}, 19(15):3875, 2002.

\bibitem{ciufolini2003gravitomagnetic}
Ciufolini, I., Kopeikin, S., Mashhoon, B. \& Ricci, F.
\newblock On the gravitomagnetic time delay.
\newblock {\em Phys. Lett. A}, 308(2--3):101--109, 2003.

\end{thebibliography}
\end{document}